\documentclass{elsarticle}

\usepackage{lineno,hyperref}

\journal{Journal of \LaTeX\ Templates}

\usepackage{graphicx}

\usepackage{amssymb}
\usepackage{latexsym}
\usepackage{amsmath}

\begin{document}

\begin{frontmatter}

\title{
Longitudinal-field fidelity susceptibility analysis of
the $J_1$-$J_2$ transverse-field Ising model
around $J_2/J_1 \approx 0.5$
}

\author{Yoshihiro Nishiyama}
\address{Department of Physics, Faculty of Science,
Okayama University, Okayama 700-8530, Japan}




\begin{abstract}

The square-lattice $J_1$-$J_2$ transverse-field (TF) Ising model
was investigated with the
exact diagonalization
(ED) method.
In order to
analyze the TF-driven phase transition,
we applied
the longitudinal-field fidelity susceptibility $\chi^{(h)}_F$, which is readily evaluated via the ED scheme.
Here,
the longitudinal field couples with 
the absolute value of the magnetic moment $|M|$ rather than the raw $M$
so that the remedied fidelity susceptibility exhibits a peak around the critical point;
note that
the spontaneous magnetization does not appear for the finite-size systems.
As a preliminary survey,
the modified fidelity susceptibility $\chi^{(h)}_F$ is applied to the analysis
of criticality for $J_2=0$, where a number of preceding results are available.
Thereby, properly scaling the distance from the multi-criticality, $\eta=0.5-J_2$,
the $\chi^{(h)}_F$ data were cast into the crossover-scaling formula,
and the multi-critical exponent for $\chi_F^{(h)}$ is estimated.
The result is cross-checked by the numerically evaluated $\beta$-function behavior.

\end{abstract}


\end{frontmatter}


\section{\label{section1}Introduction}

The three-dimensional classical Ising model
with the competing next-nearest-neighbor interaction, 
the so-called $J_1$-$J_2$ model, has been studied extensively \cite{Anjos07,Salmon13}.
The sufficiently strong frustration 
$J_2/J_1>0.5$
induces
the spatially modulated order, namely,
the stripe phase, at low temperatures,
where
slow relaxations to the thermal equilibrium
were observed even for such uniform system \cite{Hellmann93}.
Meanwhile,
the quantum counterpart, namely, the two-dimensional transverse-field 
Ising model, has come under thorough investigation \cite{Kellermann19,Bobak18}, 
shedding light on the character of the stripe phase at low temperatures
\cite{Dominguez21,Guerrero20,Jin13,Schumm24}.
In contrast, little attention has been paid to 
the moderate frustration regime
$J_2/J_1 \to 0.5^-$ at the ground state \cite{Oitmaa20,Nishiyama07,Sadrzadeh16}.
As a reference, we also mention the case of
the two-dimensional classical model.
There is no finite-temperature phase transition
at the fully frustrated point for the Villain 
\cite{Katzgraber08} 
and related 
\cite{Stephenson70,Selke09,Assis17}
models.
The similar conclusion was obtained for the $J_1$-$J_2$ model 
\cite{Murtazaev14}
as well,
although the characters around the fully-frustrated point are not 
completely understood
\cite{Lee10,Ohare09}.   
We do not pursue this issue,
because the present concern lies in the quantum model at the ground state.

In order to detect the ground-state phase transition, we employ the fidelity.
The fidelity is given by the overlap between the ground states
\cite{Uhlmann76,Jozsa94,Peres84,Gorin06}
\begin{equation}
\label{fidelity}
F(H,H+\Delta H)= |\langle H | H+\Delta H \rangle|
,
\end{equation}
with the proximate interaction parameters, $H$ and $H+\Delta H$.
Here, we choose the interaction parameter as the longitudinal 
(symmetry breaking)
field 
$H$ 
\cite{Nishiyama13,Rossini18,Bonfim19}
rather than the temperature-like (symmetry preserving) parameter 
\cite{Quan06,Zanardi06,HQZhou08,Yu09,You11,Mukherjee11}.
Restricting ourselves to $H=0$, 
the longitudinal-field fidelity susceptibility 
\cite{Nishiyama13,Rossini18,Bonfim19}
is calculated via
\begin{equation}
\label{fidelity_susceptibility}
\chi_F^{(h)} = -\frac{1}{N}
\partial^2_{\Delta H} F(0,\Delta H)|_{\Delta H=0}
,
\end{equation}
with the system size $N$.
Unlike the temperature-like fidelity susceptibility,
the
longitudinal-field fidelity susceptibility 
does not exhibit a peak around the phase transition point,
because the spontaneous symmetry breaking does not occur
for the finite-size systems
\cite{Ferrenberg91}.

In this paper, 
this flaw is remedied by replacing 
the conjugate moment $M$ of $H$ with its absolute value $|M|$
\cite{Ferrenberg91}.
It is anticipated that the modified longitudinal-field fidelity susceptibility 
exhibits a peak around the critical point;
we stress that this modification is applicable to the quantum spin model
as well.
Moreover, it 
detects the signature for the criticality sensitively,
because its scaling dimension is larger than that of the 
temperature-like fidelity susceptibility \cite{Nishiyama13}.
By means of the exact diagonalization method,
we evaluated the modified $\chi_F^{(h)}$
for 
the square-lattice $J_1$-$J_2$ transverse-field Ising model,
placing an emphasis on the multi-criticality toward the fully frustrated point
$J_2/J_1 \to 0.5^-$.
The fidelity $F$ (\ref{fidelity})
is readily evaluated with the exact diagonalization method
\cite{Yu09},
because it yields
the ground-state vector 
$| H\rangle$ explicitly.

To be specific, the Hamiltonian for the two-dimensional $J_1$-$J_2$ transverse-field Ising model
is given by
\begin{equation}
\label{Hamiltonian}
{\cal H}= -J_1\sum_{\langle ij \rangle }S^z_iS^z_j 
+J_2 \sum_{\langle \langle ij \rangle\rangle} S^z_iS^z_j
-\Gamma \sum_i^N S^x_i 
-H |M|
.
\end{equation}
Here, the spin-$S=1/2$ operator ${\bf S}_i$ 
is placed at each square-lattice point, $i=1,2,\dots,N$;
hence, the linear dimension of the cluster is given by $L=\sqrt{N}$.
The summation $\sum_{\langle  ij \rangle}$ ($\sum_{\langle\langle ij \rangle \rangle}$)
runs over all possible (next) nearest neighbor pairs 
$\langle i j \rangle $ ($\langle \langle ij \rangle\rangle$), and 
the parameter $J_1$ ($J_2$) denotes the corresponding coupling constant.
Hereafter, we consider the nearest-neighbor interaction $J_1$ as the unit of energy, 
{\it i.e.}, $J_1=1$.
The $\Gamma$ ($H$) denotes the transverse (longitudinal) field.
The longitudinal magnetic moment is given by
\begin{equation}
M=\sum_i S^z_i
.
\end{equation}
The expression $|M|$ in ${\cal H}$ (\ref{Hamiltonian})
is meant to take the diagonal value, because 
the quantization axis is parallel to the $z$ direction.
The longitudinal field $H$ is an infinitesimal perturbation, and it is irrelevant to the
physical properties such as the phase diagram.

\begin{figure}
\includegraphics[width=120mm]{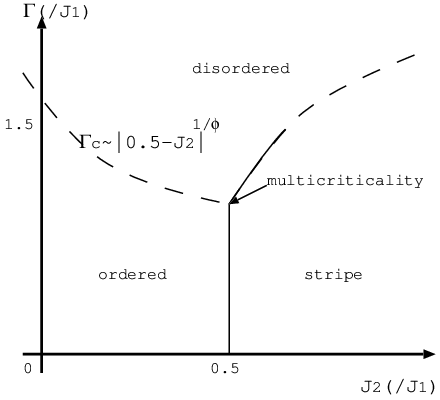}%
\caption{\label{figure1}
A schematic phase diagram \cite{Oitmaa20} for the $J_1$-$J_2$ transverse-field Ising model
(\ref{Hamiltonian})
is shown. 
The solid (dashed) line indicates the discontinuous (continuous) phase boundary.
The transverse field $\Gamma$ induces the order-disorder phase transition,
and the power-law singularity of the phase boundary,
	$\Gamma_c\sim |0.5-J_2|^{1/\phi}$ (\ref{power_law_phase_boundary}) with the
crossover critical exponent $\phi$, is one of our concerns.
For exceedingly large $J_2 > 0.5$, the stripe phase appears.
}
\end{figure}

A schematic phase diagram \cite{Oitmaa20} for the transverse-field $J_1$-$J_2$
Ising model (\ref{Hamiltonian}) is presented in Fig. \ref{figure1}.
The solid (dashed) line shows the discontinuous (continuous) phase transition.
the transverse field $\Gamma$ induces the order-disorder phase transition
at $\Gamma=\Gamma_c$,
and 
the power law singularity of the critical branch \cite{Riedel69,Pfeuty74}
\begin{equation}
\label{power_law_phase_boundary}
\Gamma_c(J_2)-\Gamma_c(0)
\sim
(0.5-J_2)^{1/\phi}
,
\end{equation}
($J_2<0.5$)
with the crossover critical exponent $\phi$ is one of our main concerns.
Eventually,
for exceedingly large frustration, $J_2>0.5$,
the stripe phase is realized.
The multi-critical point and the associated multi-critical exponents at $J_2=0.5$
have been numerically studied by the
exact-diagonalization\cite{Nishiyama07},
series-expansion \cite{Oitmaa20}, and 
tensor-network \cite{Sadrzadeh16}
methods.
In the context of the Lifshitz criticality \cite{Diehl00,Shpot01},
the multi-criticality is identified as
$(d,m)=(3,2)$,
where the parameter $d$ ($m$) denotes the total (frustrated subspace's) dimensionality.
Toward the multi-critical point \cite{Diehl00,Shpot01},
the real-space and imaginary-time correlation lengths,
$\xi$ and $\xi_\tau$, respectively,
diverge {\em anisotropically},
obeying $\xi_\tau\sim \xi^{\dot{z}}$ 
with
the dynamical multi-critical exponent $\dot{z} \ne 1$.
Hence, the exact diagonalization method has an advantage in
that the infinite imaginary-time system size 
$\beta \to \infty$
(inverse temperature) is tractable, and 
only the real system size $L$ has to be scaled carefully,
as in the ordinary isotropic finite-size scaling analyses.

In fairness, it has to be mentioned that the ordinary (temperature-like)
fidelity susceptibility exhibits suppressed corrections to
finite-size scaling \cite{Yu09,Wang15},
Moreover, 
``without prior knowledge of the local order parameter''
\cite{Wang15}, the criticality can be analyzed in a systematic manner.
It is expected that the former merit would be retained for 
the longitudinal-field-mediated $\chi_F^{(h)}$ 
(\ref{fidelity_susceptibility})
as well.

The rest of this paper is organized as follows.
In Sec. \ref{section2}, 
turning off the magnetic frustration $J_2=0$ tentatively,
we make
a finite-size-scaling analysis of the $\Gamma$-driven criticality via the probe $\chi_F^{(h)}$.
Based on this preliminary survey,
we investigate the critical branch
(\ref{power_law_phase_boundary}), placing an emphasis on the multi-criticality
at $J_2 \to 0.5^-$.
In Sec. \ref{section3}, we address the summary and discussions.

\section{\label{section2}Numerical results}

In this section, we investigate the critical branch 
(\ref{power_law_phase_boundary})
for
the $J_1$-$J_2$ transverse-field Ising model 
(\ref{Hamiltonian})
via the probe $\chi_F^{(h)}$ (\ref{fidelity_susceptibility}).
We employed the exact diagonalization method \cite{Yu09}
for the cluster
with $N \le 36$ spins.  
We refer the reader to Ref. \cite{Sandvik10},
where a brief
sample code for the quantum spin model 
with
the Lanczos algorithm is shown.
The scaling formula for $\chi_F^{(h)}$
is given by \cite{Nishiyama13,Albuquerque10}
\begin{equation}
\label{scaling_formula}
\chi_F^{(h)}=  L ^{x_F}  f((\Gamma-\Gamma_c)L^{1/ \nu})
,
\end{equation}
with a non-universal scaling function $f$, the critical point $\Gamma_c$, the correlation-length critical exponent $\nu$ 
($\xi \sim |\Gamma-\Gamma_c|^{-\nu}$),
and $\chi_F^{(h)}$'s scaling dimension
\begin{equation}
\label{scaling_relation}
x_F=\gamma_F/\nu=\gamma/\nu+z=x+z .
\end{equation}
Here, the critical exponents $\gamma_F$ and $\gamma$ denote the $\chi_F^{(h)}$ critical exponent
($\chi_F^{(h)} \sim |\Gamma-\Gamma_c|^{-\gamma_F}$), and the magnetic-susceptibility
critical exponent, 
($\chi\sim |\Gamma-\Gamma_c|^{-\gamma}$), respectively.
The exponents $z$ and $x$ are  the dynamical critical exponent \cite{Albuquerque10}
and $\chi$'s scaling dimension, respectively.
As would be apparent from Eq. (\ref{scaling_relation}),
$\chi_F^{(h)}$'s scaling dimension $x_F(>x)$ 
is larger than 
$\chi$'s.
Hence, it is anticipated that
the probe $\chi_F^{(h)}$ detects the signature for the criticality more sensitively
than $\chi$.

\subsection{\label{section2_1}
Longitudinal-field fidelity susceptibility $\chi_F^{(h)}$ analysis at $J_2=0$:
Preliminary survey
}

As a preliminary survey, we investigate the transverse-field-driven phase transition
at
$J_2=0$ via the probe $\chi_F^{(h)}$
(\ref{fidelity_susceptibility}).
The criticality 
belongs \cite{Henkel84}
to the three-dimensional
(namely, $(2+1)$D)
Ising universality class,
and specifically, the critical exponents take the following
values
\cite{Hasenbusch10,Henkel84}
\begin{equation}
\label{3D-Ising_universality}
(\nu,\gamma,z)=(0.63002,1.23719
	,1)
	.
\end{equation}

\begin{figure}
\includegraphics[width=120mm]{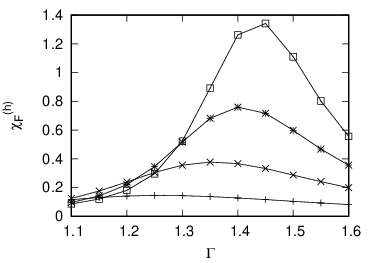}%
\caption{\label{figure2}
The longitudinal-field fidelity susceptibility 
$\chi_F^{(h)}$	(\ref{fidelity_susceptibility})
is plotted for various values of the transverse field $\Gamma$
and the system sizes,
($+$) $L=3$,
($\times$) $4$,
($*$) $5$, and
($\Box$) $6$.
Here, the frustration is 
	tentatively
	turned off, $J_2=0$, 
where a number of preceding results are avaliable \cite{Yu09,Henkel84,Huang20}.
Owing to the replacement of the conjugate moment $M$ with  $|M|$
\cite{Ferrenberg91},
the probe $\chi_F^{(h)}$ exhibits a notable peak, which indicates
an onset of the phase transition clearly.
}
\end{figure}

In Fig. \ref{figure2}, we present $\chi_F^{(h)}$ for various values of the transverse field $\Gamma$,
and 
($+$) $L=3$,
($\times$) $4$,
($*$) $5$, and
($\Box$) $6$ with $J_2=0$ fixed.
The longitudinal-field fidelity susceptibility shows a peak around the critical point 
$\Gamma_c \approx 1.4$.
Note that
the longitudinal field $H$ couples with the absolute value of the magnetic moment $|M|$
(\ref{Hamiltonian}),
and owing to this modification,
the longitudinal-field fidelity susceptibility shows a clear signature for the criticality
\cite{Ferrenberg91}.
We stress that this modification \cite{Ferrenberg91} is applicable to the quantum spin model
as well.

\begin{figure}
\includegraphics[width=120mm]{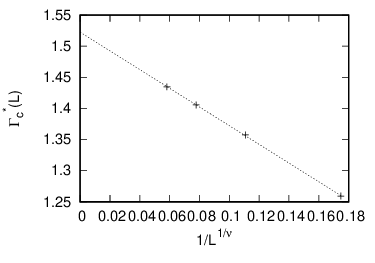}%
\caption{\label{figure3}
The approximate critical point 
$\Gamma_c^*(L)$
(\ref{approximate_critical_point})
is plotted for $1/L^{1/\nu}$ with $\nu=0.63002$ 
[Eq. (\ref{3D-Ising_universality})] and $J_2=0$.
The least-squares fit to these data yields 
an estimate
$\Gamma_c=1.5224(13)$ in the thermodynamic limit $L\to\infty$.
A possible finite-size drift error is considered in the text.
}
\end{figure}

In order to estimate the critical point precisely,
in Fig \ref{figure3}, 
we present
the approximate critical point $\Gamma_c^*(L)$ for $1/L^{1/\nu}$
with $\nu=0.63002$ 
[Eq. (\ref{3D-Ising_universality})],
$3\le L \le 6$,
and the fixed $J_2=0$.
The approximate critical point denotes $\chi_F^{(h)}$'s peak
position
\begin{equation}
\label{approximate_critical_point}
\partial_\Gamma \chi^{(h)}_F(L) |_{\Gamma=\Gamma_c^*(L)}=0
,
\end{equation}
for each system size $L$.
The abscissa scale $1/L^{1/\nu}$ is set so that the plots align,
because the expression $(\Gamma-\Gamma_c)L^{1/\nu}$ (argument of Eq. (\ref{scaling_formula}))
is dimensionless;
therefore, the critical point $\Gamma_c^*(L)$ has the power-law factor like  $\Gamma_c \sim 1/L^{1/\nu}$.
The least-squares fit to the data in Fig. \ref{figure3}
yields an estimate 
$\Gamma_c=1.5224(13)$ in the thermodynamic limit $L\to \infty$.
In order to appreciate a possible finite-size drift,
we made the same analysis as to the $L=4,5,6$ data, and arrived at an
estimate 
$\Gamma_c=1.5199(5)$; the deviation from the above one,
$\approx 2.5\cdot 10^{-3}$, appears to dominate 
the least-square-fit error $\approx 1.3\cdot 10^{-3}$.
Hence, considering the former as an indicator of uncertainty,
we estimate the critical point as
\begin{equation}
\label{critical_point}
\Gamma_c = 1.5224 (25)
.
\end{equation}
So far, for the non-frustrated case,
$J_2=0$,
a variety of analyses have been made
as to the critical point,
$\Gamma_c=1.522165(3)$ \cite{Huang20}, 
$1.525(5)$ \cite{Henkel84}, and 
$1.475(5)$ \cite{Yu09},
by means of the worm-algorithm-type
quantum Monte Carlo method,
exact diagonalization (ED)
and the ordinary-fidelity-susceptibility-mediated ED methods, respectively;
rather restricted system size $N \le 20$ was treated in Ref. \cite{Yu09},
where the methodological development of the fidelity susceptibility is focused,
and worth recollecting.
Our result $\Gamma_c=1.5224(25)$ 
[Eq. (\ref{critical_point})] appears to agree with these preceding results
\cite{Huang20,Henkel84} within the error margins,
validating the analysis via $\chi_F^{(h)}$.

\begin{figure}
\includegraphics[width=120mm]{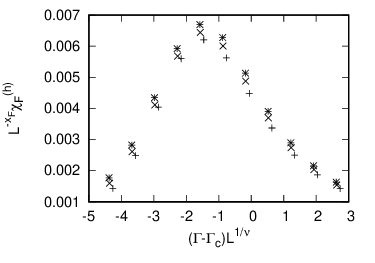}%
\caption{\label{figure4}
Based on the scaling formula (\ref{scaling_formula}),
the scaling plot,
$(\Gamma-\Gamma_c)L^{1/\nu}$-$L^{-x_F}\chi_F^{(h)}$,
	is presented for 
($+$) $L=4$
($\times$) $5$, and
($*$) $6$
with 
$\Gamma_c=1.5224$ [Eq. (\ref{critical_point})],
$\nu= 0.63002$ [Eq. (\ref{3D-Ising_universality})],
$x_F=2.96373$ [Eq. (\ref{3D-Ising_scaling_dimension})], and
$J_2=0$.
The three-dimensional Ising universality class 
(\ref{3D-Ising_universality})
is supported.
}
\end{figure}

We turn to the analysis of criticality, aiming to
examine whether the singularity belongs to the three-dimensional
(3D)
Ising universality class as mentioned in Eq. (\ref{3D-Ising_universality}).
Putting the 3D-Ising critical exponents 
(\ref{3D-Ising_universality}) 
into the scaling relation  (\ref{scaling_relation}),
we obtain $\chi_F^{(h)}$'s scaling dimension
\begin{equation}
\label{3D-Ising_scaling_dimension}
x_F = 2   .96373
.
\end{equation}
Thereby,
based on the scaling formula 
(\ref{scaling_formula}),
in Fig. \ref{figure4},
we present the scaling plot, 
$(\Gamma-\Gamma_c)L^{1/\nu}$-$L^{-x_F}  \chi_F^{(h)}$, for 
($+$) $L=4$
($\times$) $5$, and
($*$) $6$ with 
$\Gamma_c=1.5224$ [Eq. (\ref{critical_point})],
$\nu= 0.63002$ [Eq. (\ref{3D-Ising_universality})],
$x_F=2.96373$ [Eq. (\ref{3D-Ising_scaling_dimension})], and
$J_2=0$.
The scaled data appear to collapse into a scaling curve satisfactorily,
confirming that the criticality belongs to the three-dimensional Ising universality class
\cite{Henkel84}.
Note that there are no {\it ad hoc} adjustable parameters in the scaling analysis.
In fact,
the scaling parameters are all fixed in prior to the analysis,
and accordingly, the scaled data points are plotted as it is.

We address a number of remarks.
First, by means of the ordinary (temperature-like)
fidelity susceptibility, the transition point $\Gamma_c=1.53(1)$ was obtained \cite{Nishiyama19}
for the same system size as the present one.
The result appears to be consistent with ours $\Gamma_c=1.5224(25)$ 
(\ref{critical_point}) via $\chi_F^{(h)}$.
We stress that the latter $\chi_F^{(h)}$ result shows smaller error margin than the former
one.
The present $\chi_F^{(h)}$
analysis is capable of detecting the susceptibility critical
exponent $\gamma=1.23719$ 
(\ref{3D-Ising_universality}), as would be apparent from the scaling formula 
(\ref{scaling_formula})
and the relation (\ref{scaling_relation}).
Here, we emphasize that
the replacement of the conjugate moment $M$ of $H$ with $|M|$ is vital,
because the $M$-based longitudinal field fidelity susceptibility does not exhibit any peak \cite{Nishiyama13}.
To compensate this flaw, there have to be required other probes such as the Binder parameter \cite{Nishiyama13b},
which is not so reliable as mentioned below, and the consistency of the analysis becomes obscure.
Second,
the Binder-parameter and fidelity-susceptibility results were compared 
in Ref. \cite{Nishiyama13b}, and it turned out that the latter approach yields
the critical point less affected by the finite-size artifact.
Last,
the 
entanglement-mediated tensor-network simulations yield
the critical-point estimates,
$\Gamma_c=1.63$ \cite{Braiorr-Orrs16} 
and $1.64$ \cite{Shi16},
claiming that the simulation ``produces
numerical results significantly faster than QMC calculations'' \cite{Braiorr-Orrs16}.
The entanglement is an analog of the entropy, and its singularity
is basically governed by the specific-heat critical exponent \cite{Giampaolo14},
which is smaller than that of the longitudinal field susceptibility.

\subsection{\label{section2_2}
Longitudinal-field fidelity susceptibility $\chi_F^{(h)}$ analysis around 
$J_2 \to 0.5^-$
}

In this section,
we analyze the multi-criticality at $J_2=0.5$ via the probe 
$\chi_F^{(h)}$ (\ref{fidelity_susceptibility}).
For that purpose, we introduce yet another scaling parameter, $\eta=0.5-J_2$, namely,
the distance from the multi-criticality, 
and the scaling formula (\ref{scaling_formula})
is extended to 
\cite{Riedel69,Pfeuty74}
\begin{equation}
\label{multi-critical_scaling_formula}
\chi_F^{(h)} =L^{\dot{x}_F}
g
\left( (\Gamma-\Gamma_c(J_2))L^{1/\dot{\nu}}  ,   \eta L^{\phi/\dot{\nu}}  
\right),
\end{equation}
with a non-universal scaling function $g$, the critical point $\Gamma_c(J_2)$ for each $J_2$,
the correlation-length critical exponent $\dot{\nu}$ at the multi-critical point $J_2=0.5$,
and
the crossover critical exponent $\phi$ (\ref{power_law_phase_boundary}).
Here, the exponent $\dot{x}_F$ denotes $\chi_F^{(h)}$'s scaling dimension at $J_2=0.5$.
As in Eq. (\ref{scaling_relation}), this multi-critical scaling dimension satisfies
\begin{equation}
\label{multi-critical_scaling_relation}
\dot{x}_F=\dot{\gamma}_F/\dot{\nu}=
\dot{\gamma}/\dot{\nu}+\dot{z}
=\dot{x}+\dot{z}
.
\end{equation}
Here, the symbols, $\dot{\gamma}_F$ and $\dot{\gamma}$,
denote the multi-critical exponents for $\chi_F^{(h)}$ and magnetic susceptibility $\chi$, respectively.
Likewise,
the indices, $\dot{z}$ and $\dot{x}$, are  the multi-critical dynamical critical
exponent and magnetic-susceptibility's scaling dimension, respectively.

Before commencing the scaling analysis based on
Eq.(\ref{multi-critical_scaling_formula}),
we fix the multi-critical exponents 
\begin{equation}
\label{crossover_critical_exponent}
\phi=0.7
,
\end{equation}
and 
\begin{equation}
\label{multi-critical_correlation_length_critical_exponent}
\dot{\nu}=0.45
,
\end{equation}
through referring to Ref. \cite{Nishiyama07}.
The remaining one $\dot{x}_F$ is left as an adjustable parameter.

\begin{figure}
\includegraphics[width=120mm]{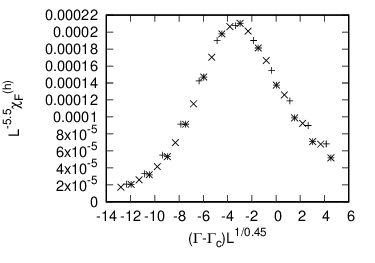}%
\caption{\label{figure5}
Based on the crossover-scaling formula 
(\ref{multi-critical_scaling_formula}),
the scaling plot,
$(\Gamma-\Gamma_c)L^{1/\dot{\nu}}$-$L^{-\dot{x}_F}\chi_F^{(h)}$,
is shown with
an optimal
$\dot{x}_F=5.5$
	[Eq. (\ref{multi-critical_scaling_dimension_value})], and
($+$) $L=4$,
($\times$) $5$, and
($*$) $6$.
The second argument of the scaling formula 
(\ref{multi-critical_scaling_formula})
is fixed to 
	$(0.5-J_2)L^{\phi/\dot{\nu}}=4$.
Here, the multi-critical exponents are set to 
$\phi=0.7$
[Eq. (\ref{crossover_critical_exponent})]
and
$\dot{\nu}=0.45$
[Eq. (\ref{multi-critical_correlation_length_critical_exponent})].
}
\end{figure}

Based on the extended scaling formula (\ref{multi-critical_scaling_formula}),
in Fig, \ref{figure5},
we present the crossover-scaling plot, 
$(\Gamma-\Gamma_c(J_2))L^{1/\dot{\nu}}$-$L^{-\dot{x}_F} \chi_F^{(h)}$,
for 
($+$) $L=4$,
($\times$) $5$, and
($*$) $6$
with 
$\dot{\nu}=0.45$ [Eq. (\ref{multi-critical_correlation_length_critical_exponent})],
an optimal $\dot{x}_F=5.5$,
and the critical point $\Gamma_c(J_2)$
determined by the same scheme as that of Sec. \ref{section2_1}.
The second argument of Eq. (\ref{multi-critical_scaling_formula}) is fixed to
$\eta L^{\phi/\dot{\nu}}=4$ with $\phi=0.7$ 
[Eq. (\ref{crossover_critical_exponent})] and 
$\dot{\nu}=0.45$ 
[Eq. (\ref{multi-critical_correlation_length_critical_exponent})].
The crossover-scaled data in Fig. \ref{figure5} 
collapse into a scaling curve satisfactorily,
confirming the validity of the scaling parameters undertaken in the analysis.
Performing the same analysis as that of Fig. \ref{figure5},
we found that
the multi-critical scaling dimension for $\chi_F^{(h)}$
lies within
\begin{equation}
\label{multi-critical_scaling_dimension_value}
	\dot{x}_F= 5.5 (3)
.
\end{equation}
As mentioned above, there is only one parameter $\dot{x}_F$
that has to be fixed
so as
to attain a good collapse of the scaled data.
Therefore, fairly straightforwardly,
the data collapse was achieved by
targeting the largest two system sizes,
$L=5$ and $6$, so that the scaled data overlap each other around the peak position.
Because the exact diagonalization data are free from the 
statistical error, 
the local-linearity measure \cite{Kawashima93}, for instance,
is not required as an indicator.
The value (\ref{multi-critical_scaling_dimension_value})
immediately
yields
the multi-critical magnetic susceptibility exponent
\begin{equation}
\label{multi-critical_susceptibility_critical_exponent}
\dot{\gamma}=1.44 (37)
,
\end{equation}
through the formula
(\ref{multi-critical_scaling_relation}),
and
$(\dot{\nu},\dot{z})=[0.45(10),0.7(2)]$ \cite{Nishiyama07}.

In Table \ref{table},
we present
the estimate $\dot{\gamma}=1.44 (37)$ 
[Eq. (\ref{multi-critical_susceptibility_critical_exponent})]
for a comparison.
So far, various analyses such as the 
$\epsilon$-expansion with and without the
Pad\'e approximation \cite{Shpot01},
exact diagonalization (ED) \cite{Nishiyama07}, and
tensor-network (TN) \cite{Sadrzadeh16}
studies 
have been made.
The present result 
supports the $\epsilon$-expansion
result 
$\dot{\gamma}=1.558$ \cite{Shpot01}
without the Pad\'e approximation,
namely, the Lifshitz criticality scenario.
On the one hand, the result \cite{Sadrzadeh16} lies out of the error margin.
In fairness, it has to be mentioned that
this novel criticality 
is supported by the sophisticated
TN analysis for the checkerboard $J_1$-$J_2$ transverse-field Ising model
\cite{Sadrzadeh19}. 
To the best of author's knowledge, it is not fully clarified whether the criticality
extends to the homogeneous $J_1$-$J_2$ model (\ref{Hamiltonian}).


\begin{table}[t]
\centering
\begin{tabular}{l l l l l l}
	method & $\Gamma_c|_{J_2=0.5}$ & $\dot{\nu}$ & $\dot{z}$ & $\phi$ &  $\dot{\gamma}$   \\ 
\noalign{\smallskip}\hline\noalign{\smallskip}
	$\epsilon$ exp \cite{Shpot01} & & $0.387$ &  $2.054$ & $0.686$ &   $1.558$ \\
	$\epsilon$ exp ($[1/1]$ Pad\'e) \cite{Shpot01} & &  $0.482$ & $2.552$  & $0.688$  & $2.02$ \\	
ED \cite{Nishiyama07} & & $0.45(10)$ & $2.3(3)$ & $0.7(2)$ & \\
	TN \cite{Sadrzadeh16} & $0.50$ & $1.0$ &  & & $0.33$ \\
series exp \cite{Oitmaa20} & $\approx 0.90$ & & & & \\
	ED	(this work) & $0.67(15)$ &  & & & $1.44(37)$ \\
\end{tabular}
\caption{
The multi-critical point $\Gamma_c|_{J_2=0.5}$, and the multi-critical exponents,
$\dot{\nu}$, $\dot{z}$, $\phi$, and $\dot{\gamma}$, have been investigated 
by means of the 
$\epsilon$-expansion 
\cite{Shpot01}
with and without the Pad{\'e} approximation,
exact diagonalization (ED) \cite{Nishiyama07},
tensor network (TN)\cite{Sadrzadeh16},
and series-expansion \cite{Oitmaa20} methods.
The transition point $\Gamma_c \approx 0.90$ is read off from Fig. 1 of Ref. \cite{Oitmaa20}.
	}\label{table}
\end{table}

Last, we address a remark.
The underlying physics behind 
the crossover-scaling plot,
Fig. \ref{figure5}, is
by no means identical to
the scaling plot,
Fig. \ref{figure4}.
In fact, the former scaling dimension,
$\dot{x}_F=5.5$ 
[Eq. (\ref{multi-critical_scaling_dimension_value})],
is substantially larger than the latter,
$x_F = 2.96373$ 
[Eq. (\ref{3D-Ising_scaling_dimension})].
Therefore, the data collapse of Fig. \ref{figure5} is by no means  accidental.

\subsection{
\label{section2_3}
Transverse-field-fidelity-susceptibility-mediated 
$\beta$-function analysis
around 
$J_2 \to 0.5^-$
}

From $\chi_F^{(h)}$ (\ref{fidelity_susceptibility}), 
we are able to extract information on the renormalization-group flow,
namely, the $\beta$ function.
The $\beta$ function indicates the derivative of the {\em effective} coupling constant
$\Gamma$ 
with respect to the concerned energy scale.
Because the $\beta$ function should exhibit a universal asymptote,
the underlying criticality is elucidated clearly.
In this section,
the multi-critical behavior found in Sec. \ref{section2_2} will be cross-checked
by the $\beta$ function analysis.
We also calculate the $\beta$ function with use of the magnetic susceptibility $\chi$,
and a comparison is made as to the scaling behavior of each probe.

The $\chi_F^{(h)}$-mediated $\beta$ function is evaluated via the expression
\cite{Roomany80}
\begin{equation}
\label{beta_function_fidelity_susceptibility}
        \beta^{(\chi_F^{(h)})}(\Gamma,L)=
        \frac{
        \dot{x}_F-
        \log(\chi_F^{(h)}(L)/\chi_F^{(h)}(L-1))/\log(\sqrt{L/(L-1)})
}{
        \sqrt{
\partial_\Gamma \chi_F^{(h)}(L) 
\partial_\Gamma \chi_F^{(h)}(L-1)/\chi_F^{(h)}(L) /\chi_F^{(h)}(L-1)
                }
                ,
}
\end{equation}
with the fidelity susceptibility $\chi_F^{(h)}(L)$ (\ref{fidelity_susceptibility})
for the system size $L$, 
and our result $\dot{x}_F=5.5$ 
[Eq. (\ref{multi-critical_scaling_dimension_value})]
for $\chi_F^{(h)}$'s scaling dimension is fed into this
formula (\ref{beta_function_fidelity_susceptibility}).
Likewise,
based on the magnetic susceptibility $\chi(L)$,
we also evaluated
\begin{equation}
\label{beta_function_susceptibility}
        \beta^{(\chi)}(\Gamma,L)=
        \frac{
        \dot{x}-
        \log(\chi(L)/\chi(L-1))/\log(\sqrt{L/(L-1)})
}{
        \sqrt{
\partial_\Gamma \chi(L) 
\partial_\Gamma \chi(L-1)/\chi(L) /\chi(L-1)
                }
                ,
}
\end{equation}
with $\chi$'s scaling dimension
\begin{equation}
	\label{multi-critical_susceptibility_scaling_dimension}
\dot{x}=3.2
  ,
\end{equation}
obtained from the scaling relation (\ref{multi-critical_scaling_relation}),
$\dot{x}_F=5.5$ 
[Eq. (\ref{multi-critical_scaling_dimension_value})],
and
$\dot{z}=2.3$ \cite{Nishiyama07}.
The $\beta$ function exhibits a universal asymptote  \cite{Roomany80}
\begin{equation}
        \label{beta_function_asymptotic_form}
        \beta^{(\chi_F^{(h)}),(\chi)} (\Gamma) = \frac{1}{\dot{\nu}}  (\Gamma-\Gamma_c)
        ,
\end{equation}
with the slope $1/\dot{\nu}$,
and the critical point $\Gamma_c$.
Because the $\beta$ function exhibits such a universal character,
the behaviors of
$\beta^{(\chi_F^{(h)})}$ (\ref{beta_function_fidelity_susceptibility}) and
$\beta^{(\chi)}$ (\ref{beta_function_susceptibility})
are compared on an equal footing.
Namely,
the deviation of $\beta^{(\chi_F^{(h)}),(\chi)}$ from the asymptote
(\ref{beta_function_asymptotic_form})
indicates an amount of  corrections to finite-size scaling.

In order to detect the multi-criticality properly,
we rely on the scaling formula (\ref{multi-critical_scaling_formula}),
for which the parameters have to satisfy
the scaling relation
\begin{equation}
\label{scaling_trajectory}
	(0.5-J_2)^{1/\phi}/(\Gamma-\Gamma_c)=0.2
.
\end{equation}
That is,
the second argument of the scaling formula
(\ref{multi-critical_scaling_formula}),
$(0.5-J_2)L^{\phi/\dot{\nu}}$,
is supposed to take
a constant value (scale invariant),
and 
through
the 
definition of $\dot{\nu}$,
{\it i.e.},
$L(\sim \xi)\sim |\Gamma-\Gamma_c|^{-\dot{\nu}}$,
the above scaling relation
(\ref{scaling_trajectory})
follows.

\begin{figure}
\includegraphics[width=120mm]{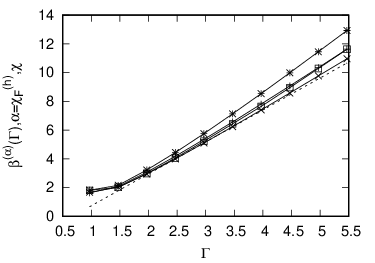}%
\caption{\label{figure6}
The $\beta$ function, $\beta^{(\alpha)}(\Gamma)$
($\alpha=\chi_F^{(h)}$ (\ref{beta_function_fidelity_susceptibility}),
$\chi$ (\ref{beta_function_susceptibility})),
is plotted for various $\Gamma$
with 
($+$) $(\alpha,L)=(\chi_F^{(h)},5)$,
($\times$) $(\chi_F^{(h)},6)$,
($*$) $(\chi,5)$, and
($\Box$) $(\chi,6)$.
By the dotted line,
the asymptote 
$\frac{1}{\dot{\nu}}(\Gamma-\Gamma_c)$
(\ref{beta_function_asymptotic_form})
with 
$\dot{\nu}=0.45$
[Eq. (\ref{multi-critical_correlation_length_critical_exponent})]
and
$\Gamma_c=0.67$ 
[Eq. (\ref{multi-critical_point})]
is shown.
}
\end{figure}

In Fig. \ref{figure6},
the $\beta$ function, 
$\beta^{(\alpha)} (\Gamma)$,
is plotted for various $\Gamma$ and
($+$) $(\alpha,L)=(\chi_F^{(h)},5)$,
($\times$) $(\chi_F^{(h)},6)$,
($*$) $(\chi,5)$, and
($\Box$) $(\chi,6)$.
The asymptote 
(\ref{beta_function_asymptotic_form})
with 
$\dot{\nu}=0.45$
[Eq. (\ref{multi-critical_correlation_length_critical_exponent})]
and
an optimal
$\Gamma_c=0.67$  is also presented by the dotted line.
The $\beta$ function obeys the asymptote for a rather
wide range of $\Gamma$.
Surveying various values of $\Gamma_c$,
we found that the multi-critical point 
lies within
\begin{equation}
\label{multi-critical_point}
\Gamma_c=0.67(15)
.
\end{equation}
This estimate
is presented in Table \ref{table};
the result $\Gamma_c \approx 0.90$
was read off from Fig. 1 of Ref. \cite{Oitmaa20}.
The present result (\ref{multi-critical_point})
locates in the middle of the preceding ones,
$\Gamma_c=0.50$
\cite{Sadrzadeh16} and
$\approx 0.90$
\cite{Oitmaa20}, determined by the TN and series-expansion methods, respectively;
the former TN result
$\Gamma_c=0.50$ appears
to be
slightly out of the error margin of ours.

A few remarks are in order.
First, the $\chi_F^{(h)}$-mediated $\beta$ function
(\ref{beta_function_fidelity_susceptibility})
obeys the anticipated asymptote 
(\ref{beta_function_asymptotic_form})
for wider range of $\Gamma$
than the $\chi$-mediated one (\ref{beta_function_susceptibility}).
Such a feature may be due to the large scaling dimension
$\dot{x}_F=5.5$ 
[Eq. (\ref{multi-critical_scaling_dimension_value})] of 
the former, $\chi_F^{(h)}$.
Therefore, the singular part of $\chi_F^{(h)}$
may
dominate
the scaling corrections,
such as the 
regular 
(non-singular) 
part, around the critical point.
Last, 
the analysis in Sec. \ref{section2_2} is cross-checked by the $\beta$-function behavior
(\ref{beta_function_asymptotic_form}).
Actually
we fixed 
the multi-critical exponents, 
$\dot{x}_F$ 
(\ref{multi-critical_scaling_dimension_value}),
$\dot{x}$ 
(\ref{multi-critical_susceptibility_scaling_dimension}),
$\phi$  
(\ref{crossover_critical_exponent}), and
$\dot{\nu}$ 
(\ref{multi-critical_correlation_length_critical_exponent}),
in prior to the analysis,
and no
{\it ad hoc} adjustable parameters
are undertaken in the analysis of this section.

\section{\label{section3}
Summary and Discussions}

The $J_1$-$J_2$ transverse-field Ising model 
(\ref{Hamiltonian}) was investigated numerically.
The end-point singularity of the critical branch
(\ref{power_law_phase_boundary}),
namely, the multi-criticality,
is our main concern.
As a probe to detect the phase transition,
we utilized the longitudinal-field fidelity susceptibility 
$\chi_F^{(h)}$ (\ref{fidelity_susceptibility}),
replacing the conjugate moment $M$ of $H$ with its absolute value
$|M|$ 
\cite{Ferrenberg91} 
so as to realize the peak around $\Gamma=\Gamma_c$.
As a preliminary survey, the probe $\chi_F^{(h)}$
was applied to the $J_2=0$ case, where preceding results are available
\cite{Henkel84,Huang20,Yu09}.
Thereby,
it turned out that 
these results are reproduced by
the $\chi_F^{(h)}$-mediated scheme.
We then turn to the analysis of the multi-criticality at $J_2 \to 0.5^-$
by extending the scaling formula (\ref{scaling_formula}) to include yet another
scaling parameter, $\eta=0.5-J_2$.
The crossover-scaled $\chi_F^{(h)}$
data yield the multi-critical magnetic susceptibility exponent
$\dot{\gamma}=1.44 (37)$ 
[Eq. (\ref{multi-critical_susceptibility_critical_exponent})].
As a cross-check,
the $\chi_F^{(h)}$-mediated $\beta$-function (\ref{beta_function_fidelity_susceptibility})
was evaluated, 
confirming 
the consistency of our scheme.
As a byproduct, the critical point 
$\Gamma_c=0.67(15)$
[Eq. (\ref{multi-critical_point})]
was estimated by the zero point of the $\beta$ function.
As summarized in Table \ref{table}, the present results support the $\epsilon$-expansion analysis
\cite{Shpot01}
based on the Lifshitz criticality scenario.

In Ref. \cite{Schmitt22},
various approaches,
{\it i.e.}, the
infinite-projected-entangled-pair, matrix-product,
and
neural-quantum states,
are compared, claiming that these approaches have
"complementary regimes of applicability" \cite{Schmitt22}.
The present method is applicable to
the off-multi-critical regime $\eta>0$, as shown in Sec. \ref{section2_1},
and these $\eta>0$ data were targeted and cast into the crossover-scaling formula (\ref{multi-critical_scaling_formula}).
We owe the idea to 
Ref. \cite{Mukherjee11},
where the multi-criticality of the one-dimensional $XY$ model
is analyzed
with the ordinary fidelity susceptibility,
claiming that the properly scaled trajectory toward the multi-critical point
captures the multi-criticality.
It has to be mentioned, however, that 
the present method is not applicable to the direct analysis
of the regime at the fully frustrated point,
where intriguing stripe-pattern fluctuations were found \cite{Jin13,Schumm24}.

Another criticality scenario
is advocated
for the checkerboard-lattice $J_1$-$J_2$ transverse-field Ising model \cite{Sadrzadeh19},
for which a sophisticated tensor renormalization group method was developed.
It would be tempting to consider the 
transient behavior
between the checkerboard and homogeneous $J_1$-$J_2$ models.
This problem is left for the future study.

\section*{Acknowledgment}

This work was supported by a Grant-in-Aid
for Scientific Research (C)
from Japan Society for the Promotion of Science
(Grant No.
20K03767).

\section*{References}


\end{document}